\pdfoutput=1 %arXiv admin added this line
\documentclass[a4paper,10pt,twoside]{cpc-hepnp}

\usepackage{multicol}
\usepackage{graphicx}
\usepackage{booktabs}
\usepackage{amssymb,bm,mathrsfs,bbm,amscd}
\usepackage[tbtags]{amsmath}
\usepackage{lastpage}

\begin{document}

\fancyhead[co]{\footnotesize M. Ajaz~ et al: Instruction for typesetting manuscripts}

\footnotetext[0]{Received 04 May 2012}

\title{Study of the behavior of the nuclear modification factor in
freeze-out state}

\author{%
      M. Ajaz$^{1,2;1)}$\email{muhammad.ajaz@cern.ch}%
\quad M. K. Suleymanov$^{1,3;2)}$\email{mais$\_$suleymanov@comsats.edu.pk}
\quad K. H. Khan$^{1}$
\quad A. Zaman$^{1}$
}
\maketitle

\address{%
$^1$ Department of Physics, COMSATS Institute of Information Technology, 44000, Islamabad, Pakistan\\
$^2$ Department of Physics, Abdul Wali Khan University, 23200, Mardan, Pakistan\\
$^3$  Veksler - Baldin Laboratory of High Energy, Joint Institute for Nuclear Research, 141980, Dubna, Russia\\
}

\begin{abstract}:
One of the latest trends in the advancement of experimental
high-energy physics is to identify the quark gluon plasma
(QGP) predicted qualitatively by quantum
chromodynamics (QCD). We discuss whether nuclear transparency effect 
which is considered an important phenomenon, connected with dynamics of 
hadron-nuclear and nuclear-nuclear interactions could reflect some particular properties of the medium. 
FASTMC is used for Au-Au collisioin at RHIC energies. Critical change in the transparency is considered 
a signal on the appearance of new phases of strongly interacting matter and the QGP.

\end{abstract}



\begin{multicols}{2}

\section{Introduction}

One of the latest trends in the advancement of experimental
high-energy physics is to search for new states of strongly interacting matter and 
to identify the quark gluon plasma (QGP)~\cite{qgp} predicted qualitatively by quantum
chromodynamics (QCD)~\cite{qcd}. QGP is considered a state of
strongly interacting matter under extreme conditions (high
temperatures and/or densities of the baryons). This can be brought
about under laboratory conditions during the collisions of relativistic
heavy nuclei by increasing the energies and varying the masses of colliding
nuclei. This leads to a continuing quest of leading research which centers
on high-energy physics to create new accelerators of heavy nuclei
and enhance the energies of  existing accelerators. Now to create the states of 
strongly interacting matter and the possible formation of a quark gluon plasma (QGP), collisions of nuclei at very
high energies and high densities are required. Since the collisions are very violent and
the time scales involved are very short, the formation of QGP depends on the initial conditions 
for the matter produced in the collision. More precisely, it depends on the distributions of partons in the wavefunctions 
of the two nuclei before the collision. Multi-particle production in high energy collisions is dominated by modes in the nuclear
wave-function carrying a small momentum fraction x of the nuclear momentum~\cite{[11]}. Understanding the correct
initial conditions therefore requires the understanding of the properties of these small x modes. 
A sophisticated effective field theory approach has been developed to describe the properties of partons at
small x ~\cite{[22], [33], [44]} forming a color glass condensate
(CGC) ~\cite{[22], [55], [66]}. The CGC is characterized by a bulk
scale $\Lambda_{s}$ which grows with energy and the size of the nuclei.
For RHIC energies, $\Lambda_{s}$ is of the order of 1-2 GeV ~\cite{[77], [88], [99], [100]} 
provided $\Lambda_{s}$ is a constant as for cylindrical nuclei.
The field strengths in the saturation region behave as about {\Large $\frac{1}{\alpha_{s}}$}: 
since $\alpha_{s}$($\Lambda_{s}$) $<<$ 1, the field strengths are large. Furthermore, the occupation number of saturated gluons
is also of the order of {\Large $\frac{1}{\alpha_{s}}$}  $>>$ 1. For more details, the interested reader is referred 
to ~\cite{[111], [112], [113]} where the initial conditions for nuclear collisions are formulated in the CGC.
Strongly interacting matter may be subject to a series of phase
transitions with increasing temperature and/or density of the baryon
among which is the first-order phase transition of restoration of a
specific symmetry of strong interactions---chiral symmetry that is
strongly violated at low temperatures and/or densities of the baryon
charge. However, to create necessary laboratory conditions and pick
up a "signal" of formation of the QGP phase, one needs a lot of
intellectual and material resources.  The well known time evolution
of central heavy ion collisions is shown in Fig.~\ref{figj1}. Several phases of a typical heavy ion collision can be identified.
Here the five states are shown as: I. pre-interaction state;
II. parton-parton interaction one and the mixed phase; III. QGP phase ; IV. hadranization and V. freezout.
In the pre-interaction state, two nuclei are approaching each other with high velocity. their orientation in space and the
 initial beam direction define the reaction plane. The impact parameter b is located in the reaction plane which is a perpendicular
 distance between the center of the two colliding nuclei. In the case of most central collision the value of b is equal to zero. 
In the second stage when the two heavy nuclei approach each other and start to overlap, the parton-parton interaction is expected 
because of high energy and centrality, which leads to the mixed phase of the strongly interacting system. After the overlap of the matter 
distributions of the two colliding nuclei, the properties of the nucleon-nucleon interaction are not well known. Medium effect -the modification 
of the properties of the constituents, and at high energies the transition to the QGP occurs. The critical value of temperature for the formation 
of such plasma is of the order of 170 MeV. The idea of QGP is illustrated with the following example. Consider a certain volume containing baryons. 
Experiments have shown that baryons have non-zero spatial volume ~\cite{[114]}. Thus clearly a critical volume exists where baryons fill the volume 
completely and at this critical volume it is assumed that the baryon structure vanishes and forming the plasma of quarks and gluons. Here it is worth 
pointing out that the quarks are still confined by the strong interaction but now instead of being confined within hadrons they are confined within 
the allowed volume. Now the energy density is much higher in this volume unlike the environment found inside nucleons. The matter inside the 
volume is proposed to be in a state known as a QGP. It is believed that the early universe went through a QGP phase as it expanded and 
cooled. Hadronization is the next stage in the reaction setup where the relaxation of the energy density takes place. The central 
system is undergoing expansion, thereby reducing its temperature and density. For symmetry reasons, the expansion is 
azimuthally symmetric for central collisions. The matter begins to expand collectively into transverse directions. The features 
of this process may be understood at a qualitative level in terms of the self-similar cylindrically symmetric hydrodynamic expansion. 
In the self-similar expansion, the velocity is proportional to the distance from symmetry axis. The system will expand and cool until the 
energy densities and temperatures are no longer sufficient to sustain a thermally equilibriated system. At a temperature of about 15 fm/c, 
the momenta of the particle will kinetically “freeze-out” and the particles will stream freely without further strong interactions. In each state, 
the matter can be characterized by different temperature and density. Apart from these parameters, there could be another very interesting 
parameter, namely the transparency ($Tr$) of matter, to characterize these states. We believe that the appearance and changing of the 
transparency could give the necessary information for the identification of QGP formation. To extract the nuclear transparency, 
nuclear modification factor $(R)$ is used. In the literature one can find two different ways of defining nuclear modification 
factor denoted by $R_{AA}$ and $R_{CP}$. The former is defined as the ratio of the particles yield in nucleus-nucleus collision 
normalized to the number of binary collisions to that of nucleon-nucleon collisions, whereas the latter is defined as the ratio of the 
particles yield in the central to peripheral collisions. Sometimes, due to the lack of appropriate pp data, which enables to calculate $R_{AA}$, 
a ratio of central to peripheral spectra is used ($R_{CP}$), on the premise that ultra peripheral
events look very much like elementary collisions. We use the latter one in the current study.

\end{multicols}
\ruleup
\begin{center}
\includegraphics[width=15cm]{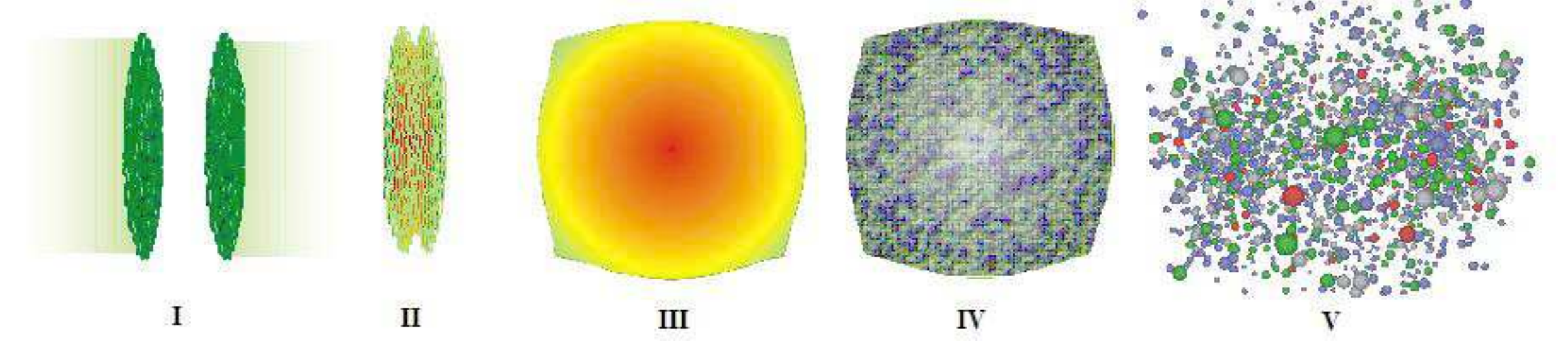}
\figcaption{\label{figj1} A schematic view of the time evolution of the central heavy ion collision. Image prepared by Steffen Bass, Duke University}
\end{center}
\ruledown

\begin{multicols}{2}

In Ref.~\cite{ssstz} a signal on mixed phase formation is considered an important point to identify  QGP because 
such a state must contain real particles which can interact with matter and therefore could be detected.

\section{The nuclear modification factor}
In this paper we discuss some ideas connected with the identification of QGP using the information coming from the freezout state. 
The main idea is that the values of transparency ($Tr$) for different
states of time evolution of heavy ion collisions are different ($Tr_{III}$, $Tr_{IV}$ and $Tr_{V}$). 
To characterize the $Tr$ it is convenient to use the nuclear
modification factor ($R$). A comparison of yields in different ion
systems by using nuclear modification factors (involving central and
peripheral collisions) should provide information on the
hadronization~\cite{[4]} (see Fig.~\ref{fig2}). $R$ highlights the
particle type dependence at intermediate $p_T$ as  was suggested by the
coalescence models~\cite{[5]} leading to the idea that hadrons
result from the coalescence of quarks in the dense medium.
Using R = {\Large $\frac{n_{1}}{n_{2}}$}  (here e.g., ${n_{1}}$ and ${n_{2}}$ could be heavy flavor particles yields with fixed values 
of $p_{T}$ and $\eta$) as a function of centrality, the masses and energy it is possible to get necessary information on the properties of 
the nuclear matter. In such definition, the appearance of transparency could be identified and detected using the condition R=1.
The data obtained by STAR RHIC BNL ~\cite{[4]} on the behavior of the nuclear modification factors of the strange particles as a 
function of the centrality in Au+Au- and p+p-collisions at 200 GeV (see Fig.2) may help us to answer the questions on how the new phases of strongly interacting matter form.
To probe the processes governing this production, yields of (anti-)strange particles have been normalized as a function of the number 
of participants $N_{part}$ (usually related to soft processes) and of the number of binary collisions $N_{bin}$ (supposed to describe hard processes) 
for Au-Au collisions as it is shown in the upper and lower frames of Figure 2, respectively. Strange particles deviate from the $N_{part}$ scaling and follow a better scaling with $N_{bin}$.
We may expect a signal on the formation of the intermediate nuclear system e.g., nuclear cluster. The strange particles could be formed as a 
result of quark coalescence in high density strongly interacting matter and on the other hand they could be captured by this system intensively. 
So by increasing the centrality, the yields of heavy flavors could decrease. The appearance of superconducting property of the 
strongly interacting matter ~\cite{[4a]} as a result of the formation of percolation cluster could stop the decrease of yields of heavy flavors. 
Thus we expect to see the new structure in the behavior of the particles yields as a function of the centrality which could connect with the percolation cluster formation.

\begin{center}
\includegraphics[width=7cm]{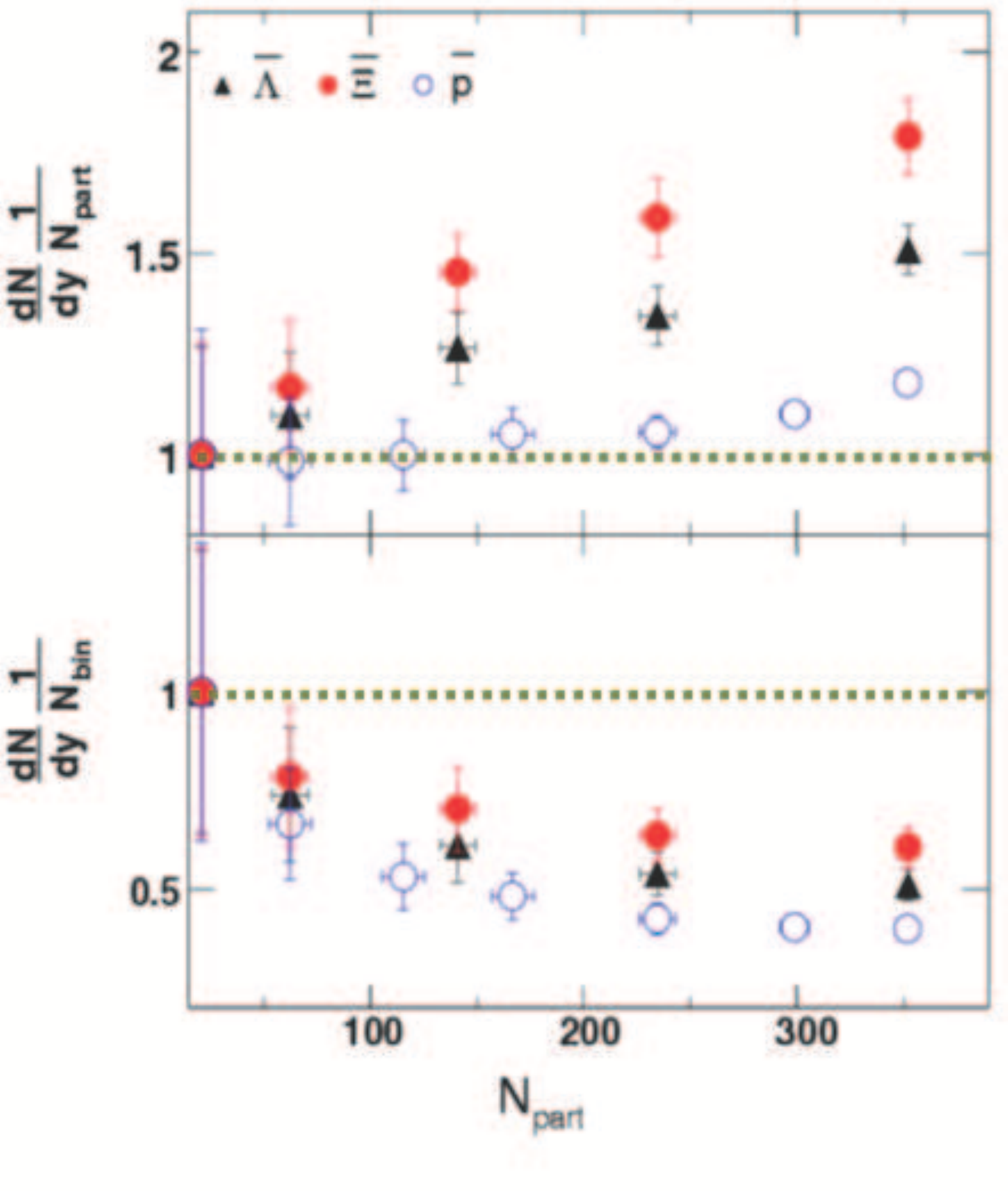}
\figcaption{\label{fig2}  Yields of some
baryons as a function of the centrality (expressed with the number
of participants) in Au+Au collisions normalized to the most
peripheral point and to number of participant $(N_{part})$ shown in the upper frame and number of binary 
collisions $(N_{bin})$ shown in the lower frame  ~\cite{[4]}. $N_{part}$ and $N_{bin}$ are calculated in the 
framework of the Glauber model. Different markers with different colors correspond to different baryons as shown.}
\end{center}
Fig.~\ref{fig3} shows the results of the PHENIX, RHIC on the studies of $J$/$\psi$ production from p+p 
to central Au+Au collisions. Results from PHENIX include $R_{AA}$ distributions for the Au$+$Au and Cu$+$Cu 
collisions ~\cite{[6]}. The data shows an increasing suppression with $N_{part}$. At central Au+Au collisions the suppression 
is larger than predicted by normal absorption in cold nuclear matter production ~\cite{[7],{rhicj}}.

\begin{center}
\includegraphics[width=7cm]{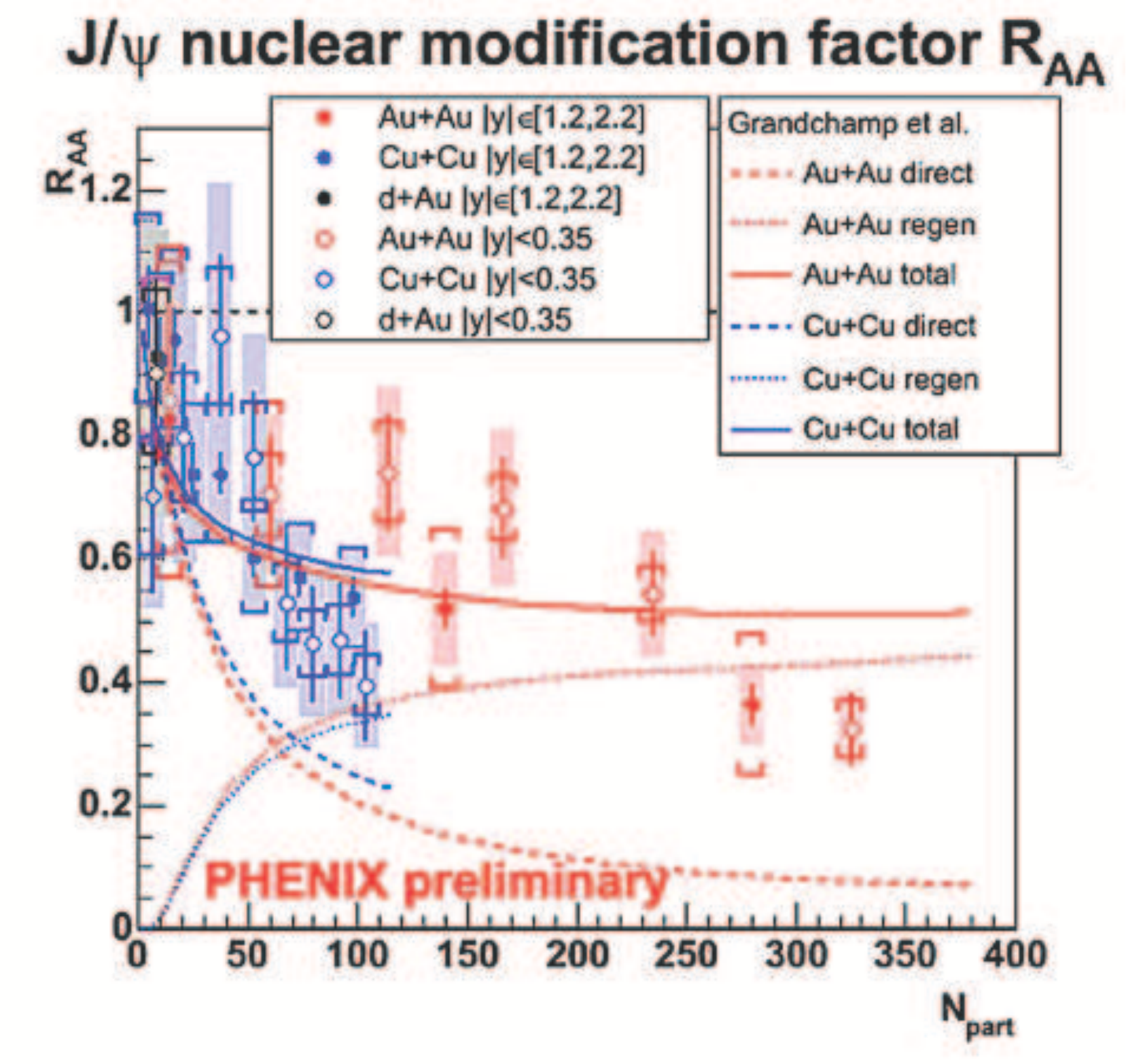}
\figcaption{\label{fig3}   Nuclear modification factor 
as a function of a number of participants ($N_{part}$) for Au+Au and Cu+Cu collisions ~\cite{[6]}. }
\end{center}
It is  supposed that $Tr_{III-V} \sim R_{III-V}$. To restore the
time scale we are going to use the values of temperature because
they must be different for III-V states.
If the states of III-Vl appear critically, the regime change will
have to be observed in the behavior of $R$ as a function of temperature.

\section{Results}
	To confirm the above idea we use data coming from different heavy
ion generators and experiments. Fig.~\ref{figj2d} shows the result of
the  study of the behavior of $R$ function defined as ratio of the
yields of different particles at central to peripheral collision as
a function of the thermal freeze-out temperature ($T_{th}$) produced
in Au-Au collisions at RHIC energies. Different colors in Fig. 4 correspond to different particles/anti-particles. 
The behavior of nuclear modification factor is studied for all the octet baryons and a few octet pseudo scalar mesons 
as well as for their corresponding anti-particles. The results show that the behavior is independent of the type of particle used. The data is simulated using
the Fast Hadron Freezout Generater (FASTMC)~\cite{[1]}. The FASTMC
hadron generation allows one to study and analyze various
observables for stable hadrons and hadron resonances produced in the
ultra-relativistic heavy ion collisions. Particles can be generated
on the chemical or thermal freeze-out hyper surface represented by a
parametrization or a numerical solution of relativistic
hydrodynamics with given initial conditions and equation of
state~\cite{[1]}. There are two regions in the behavior of  $R$ as a
function of the $T_{th}$ (see Fig.~\ref{figj2d}). In the first region
one can see that in the freeze-out state $R$ is almost a linearly
increasing  function of the $T_{th}$ independent of the types of
particles and the second region is a straight line, which has no
dependence on $T_{th}$, it can be regarded as a regime change. Now one can see that the study of the nuclear modification factor 
as a function of thermal freeze-out temperature shows a critical change and saturation at a temperature about 150 MeV. This critical change 
in $R$ is actually due to the critical change in the transparency of the medium, which shows the changing properties of the medium.

\begin{center}
\includegraphics[width=7cm]{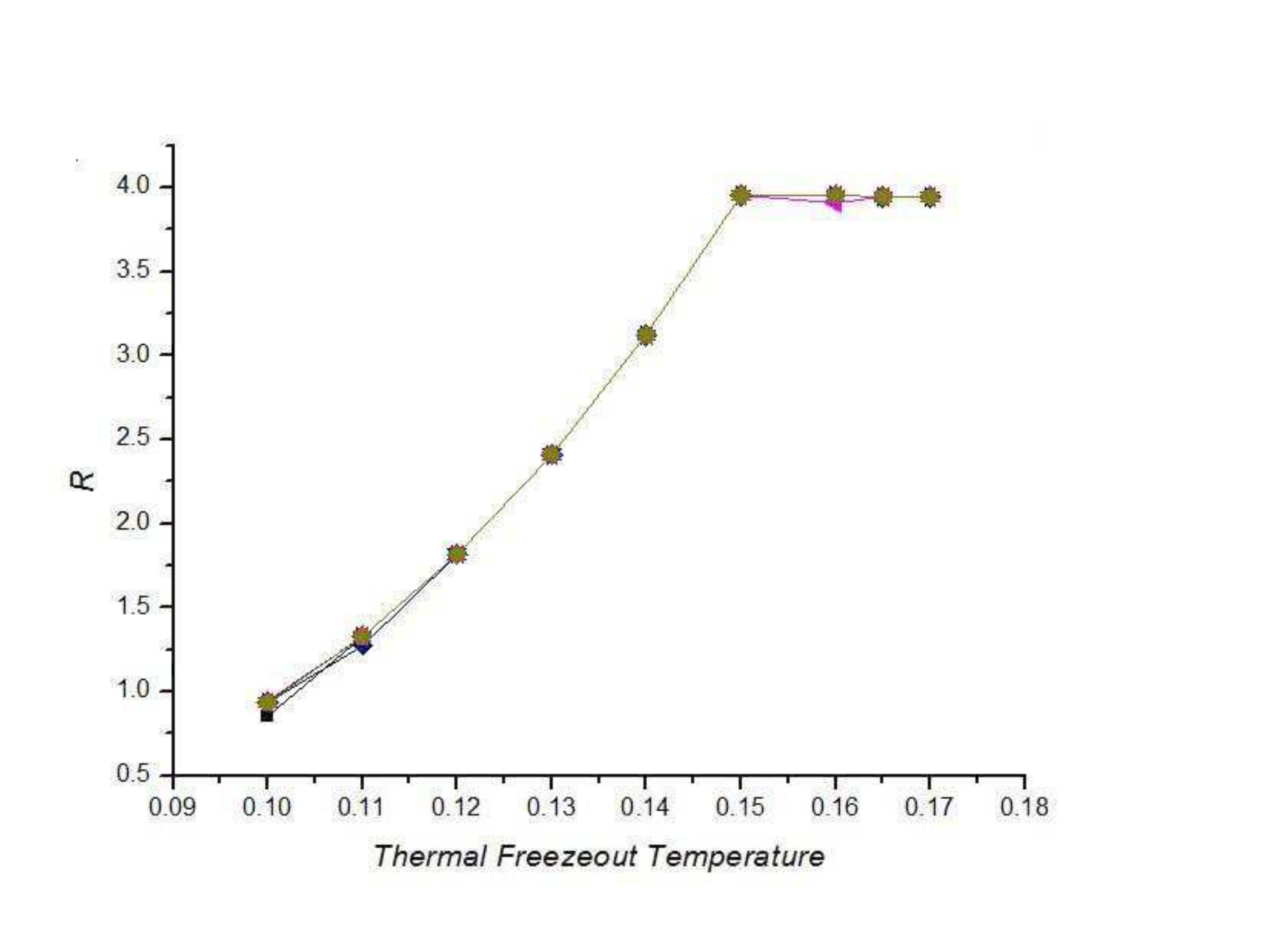}
\figcaption{\label{figj2d}  The behavior of the nuclear modification factor is shown as a function of thermal freeze-out temperature. 
Different colors in the figure correspond to the ratio of different particles/anti-particles yield produced in central Au+Au collisions to that of the 
corresponding particles/anti-particles yield in peripheral Au+Au collisions. The behavior of R is independent of the type of particle used.}
\end{center}

\section{Summary}
	We have discussed that the appearance of the critical nuclear transparency, as a function of different kinematical parameter 
of nucleus-nucleus collision is considered a signal of phase transition in nuclear mater. FASTMC is used for Au-Au collision at RHIC energies. 
This model of hadron generation allows one to study and analyze various observables for stable hadrons and hadron resonances produced in ultra-relativistic 
heavy ion collisions. The behavior of $R$ function is studied as a function of thermal freeze-out temperature. Critical change in the transparency is considered 
a signal on the appearance of new phases of strongly interacting matter and the QGP. Nuclear modification factror is used to extract the information on 
nuclear transparency effect. A critical change in the behavior of $R$ function at a temperature of about 150 MeV is observed.\\
\\

\end{multicols}
\clearpage
\end{document}